\documentclass[aps,prb,twocolumn,showpacs,amsmath,amssymb]{revtex4}
\usepackage{dcolumn}
\usepackage{bm}
\usepackage{graphicx}
\usepackage{subfigure}
\usepackage{doi}

\begin{document}
\title{Skyrmion spin texture in ferromagnetic semiconductor--superconductor heterostructures.}
\author{Kristofer Bj\"{o}rnson}
\affiliation{Department of Physics and Astronomy, Uppsala University, Box 516, S-751 20 Uppsala, Sweden}
\author{Annica M. Black-Schaffer}
\affiliation{Department of Physics and Astronomy, Uppsala University, Box 516, S-751 20 Uppsala, Sweden}
\date{\today}

\begin{abstract}
We provide a derivation of a spin Skyrmion number classification for two-dimensional topological superconductors constructed from ferromagnetic and Rashba spin-orbit coupled semiconductor-superconductor heterostructures. We show that in the non-trivial topological phase, characterized by a non-zero Chern number, there is always a topological spin texture in the occupied bands represented by a Skyrmion number. The Skyrmion number has the advantage of being both physically intuitive and directly measurable using spin-sensitive band structure imaging techniques. In addition, we show that the Skyrmion classification can be extended to the equivalent one-dimensional topological superconductors.
\end{abstract}

\maketitle

\section{Introduction}
Classification of matter into different ordered phases is a central theme in condensed matter physics. Traditionally, ordered phases have been understood using the Ginzburg-Landau paradigm,\cite{Landaubook} where spontaneous symmetry breaking gives rise to a finite order parameter, such as a finite magnetization in a ferromagnet. 
Starting with the integer quantum Hall effect,\cite{vonKlitzing80} another classification paradigm based on the topological properties of the band structure has also emerged.\cite{Thouless82, Wen95}
With the recent discovery of topological insulators\cite{Science.1133734, Science.1148047, Fu&Kane07, Hsieh08, Xia09, Zhang09, Chen09} it has now become a reoccurring task to classify different phases of matter according to the topological properties of their (single-particle) band structure.\cite{Kane&Mele05x2, Fu&Kane&Mele07, Moore&Balents07, Qi08, Roy09, RevModPhys.82.3045, RevModPhys.83.1057}

Superconductors with a full pairing gap can be classified into different topological classes in much the same way as band insulators.\cite{PhysRevB.78.195125,Qi09PRL}
Non-trivial topological superconductors have lately received an enormous amount of attention,\cite{RevModPhys.83.1057, Alicea12, Beenakker12} due to the possibility of them hosting Majorana fermions at e.g.~interfaces and vortices.\cite{Read00, Kitaev01}
The Majorana fermion is its own anti-particle and obeys non-Abelian statistics, which can be used for fault-tolerant topological quantum computation.\cite{Nayak08} The Majorana state is a consequence of the bulk-boundary correspondence, which establishes that non-trivial topological order in the bulk necessarily gives rise to boundary states crossing the bulk energy gap.\cite{RevModPhys.82.3045, RevModPhys.83.1057, PhysRevB.78.195125} In some topological superconductors the resulting zero energy modes are Majorana fermions. 
Proposals for 2D topological superconductors with Majorana fermions exist already, in e.g.~superconducting hybrid structures between a conventional $s$-wave superconductor and topological insulators\cite{Fu&Kane08,Fu&Kane09} or ferromagnetic and Rashba spin-orbit coupled semiconductors with finite magnetization.\cite{Sato09, Sato10,Sau10, Lutchyn10,Oreg10} The latter alternative is particularly interesting since semiconductor technology is very mature, there exists semiconductors with large spin-orbit coupling\cite{Yang06, Sakano13} and with demonstrated superconducting proximity effect\cite{PhysRevB.55.8457}, while ferromagnetism can be introduced either through proximity to a ferromagnetic insulator or through an applied magnetic field. All key ingredients are thus in place for finding Majorana fermions.\cite{Sato09, Sato10, Sau10, Mao10, Bjornson13} In fact, possible experimental signatures of Majorana fermions have already been reported in similar one-dimensional (1D) nanowire semiconductor structures.\cite{Mourik12, Rokhinson12, Das12}

Although topological classification is a very powerful concept, the topological invariant usually remains a rather abstract quantity.\footnote{Although examples of physically accessible topological invariants do exist in for example cold atom systems\cite{NatPhys.9.795,PhysRevA.88.013622}} The most prominent example is probably the Chern number, used for 2D time-reversal symmetry breaking systems,\cite{Thouless82, TopologicalInsulatorsAndTopologicalSuperconductors} including the 2D ferromagnetic semiconductor-superconductor heterostructures mentioned above.\cite{Sato09, Sato10}
In this article we show that the topological phase for these heterostructures instead can be classified by a sum of the Skyrmion number for the spin polarization in each band; that is, according to how the spin texture of the different bands wraps around the unit sphere, with the topological invariant given by the sum of the topological spin texture in each band.
The Skyrmion classification gives simultaneously both the topological invariant and a very intuitive physical picture of its meaning, thus making the topological order more accessible. Most importantly, the spin Skyrmion numbers can be directly measured using a spin-polarized band structure imaging technique such as spin-polarized angle-resolved photoemission spectroscopy (ARPES). This opens up for direct experimental classification of the topological phase without having to rely on indirect methods such as the existence of boundary modes. Moreover, we show why such imaging need only to be performed at certain high symmetry points, which is an important practical advantage.
We also extend the results for the 2D case to 1D wires\cite{Lutchyn10, Oreg10} for which experiments has reported Majorana fermion signatures.\cite{Mourik12, Rokhinson12, Das12} We do so through dimensional reduction of the 2D case, and show that the spins covers a circle rather than a full sphere in the topologically non-trivial phase. This provides a unified framework within which both 1D and 2D systems can be studied.

\section{From Chern to Skyrmion}
\label{Section:Chern_Skyrmion}
Ferromagnetic semiconductor-superconductor heterostructures with strong Rashba spin-orbit interaction in 2D have previously been topologically classified using a Chern number.\cite{Sato10} In general, the Chern number for a Hamiltonian $\mathcal{H}(\mathbf{k})$  is given by\cite{TopologicalInsulatorsAndTopologicalSuperconductors}
\begin{align}
\label{Equation:Chern_number}
I^{(\lambda)} &= \frac{1}{2\pi}\int dk_x dk_y\mathcal{F}^{(\lambda)}(\mathbf{k}),\\
\label{Equation:Berry_curvature}
\mathcal{F}^{(\lambda)}(\mathbf{k}) &= \epsilon^{ij}\partial_{k_i}\mathcal{A}_j^{(\lambda)}(\mathbf{k}),\\
\label{Equation:Berry_connection}
\mathcal{A}_j^{(\lambda)}(\mathbf{k}) &= -Im\left(\langle\Psi^{(\lambda)}(\mathbf{k})|\partial_{k_j}\Psi^{(\lambda)}(\mathbf{k})\rangle\right),
\end{align}
where Eq. (\ref{Equation:Chern_number}), (\ref{Equation:Berry_curvature}) and (\ref{Equation:Berry_connection}) are the Chern number, Berry curvature, and Berry connection, respectively. Here $|\Psi^{(\lambda)}(\mathbf{k})\rangle$ is the eigenstate of $\mathcal{H}(\mathbf{k})$ in band $\lambda$ at point $\mathbf{k}$ in the Brillouin zone.
It is the sum of the Chern numbers for all occupied bands that is used in actual classifications, but we will here consider the Chern number for each band individually and only add them in the end. This is in general allowed as long as separate bands, with physical properties continuous in $\mathbf{k}$, can be unambiguously defined. One exception is the topological phase transition, where the band assignment is necessarily ambiguous.

Here we will show that the Chern number defined in Eq.~(\ref{Equation:Chern_number}) for the particular case of a two-component spinor field over a 2D compact manifold can be viewed as a Skyrmion number. For all practical purposes the compact space is the Brillouin zone and the Skyrmion number simply tells us how many times the spin covers the unit sphere as it sweeps the Brillouin zone. This is a previously known result\cite{PhysRevB.74.085308} but expressed using a slightly different formulation that prepares for the results in the sections that follows.

We thus assume a single two-spinor field $|\Psi^{(\lambda)}(\mathbf{k})\rangle = e^{i\theta_{\mathbf{k}}}\left[\begin{array}{cc}u_{\uparrow\mathbf{k}}^{(\lambda)} & u_{\downarrow\mathbf{k}}^{(\lambda)} \end{array}\right]^T$ defined over the Brillouin zone, where $\uparrow(\downarrow)$ is the up (down) spin index, and an arbitrary phase factor $e^{i\theta_{\mathbf{k}}}$ has been factored out. The Berry connection in Eq. (\ref{Equation:Berry_connection}) is then given by
\begin{equation}
\label{Equation:Berry_connection_two_spinor}
\mathcal{A}_j^{(\lambda)}(\mathbf{k}) = -Im\!\!\left(\!u_{\uparrow\mathbf{k}}^{(\lambda)*}\frac{\partial u_{\uparrow\mathbf{k}}^{(\lambda)}}{\partial k_j} + u_{\downarrow\mathbf{k}}^{(\lambda)*}\frac{\partial u_{\downarrow\mathbf{k}}^{(\lambda)}}{\partial k_j} + i\frac{\partial\theta_{\mathbf{k}}}{\partial k_j}\!\right)\!.
\end{equation}
The third term disappears when calculating the Berry curvature in Eq. (\ref{Equation:Berry_curvature}) and will therefore be ignored. This allows us to redefine $\mathcal{A}_j^{(\lambda)}(\mathbf{k})$ to be the sum of the two first terms only, while at the same time assuming that $u_{\mathbf{k}\uparrow}^{(\lambda)}$ is real by letting its phase dependence be captured by $e^{i\theta_{\mathbf{k}}}$. An exception to this construction occurs at points where $u_{\mathbf{k}\uparrow}^{(\lambda)} = 0$, because at these points a unique $\theta_{\mathbf{k}}$ cannot be defined. However, at these points the Berry connection is given by an expression of the form
\begin{align*}
\mathcal{A}_j^{(\lambda)}(\mathbf{k}) =& -Im\left(|u_{\downarrow\mathbf{k}}^{(\lambda)}|\frac{\partial |u_{\downarrow\mathbf{k}}^{(\lambda)}|}{\partial k_j} + i|u_{\downarrow\mathbf{k}}|^2 \frac{\partial\theta_{\downarrow\mathbf{k}}}{\partial k_j}\right),
\end{align*}
where $\theta_{\downarrow\mathbf{k}}$ is the full complex phase of the down component. The first term in this expression gives zero contribution because it is real, while the second term cancels out in the same way as the phase factor in Eq. (\ref{Equation:Berry_connection_two_spinor}) canceled out when the Berry connection in Eq. (\ref{Equation:Berry_curvature}) is calculated. We therefore see that the contribution to the Chern number from such points are zero, and we simply chose do redefine the integration region in Eq. (\ref{Equation:Chern_number}) such that these points are excluded from the integral.

Having justified the assumption of a real $u_{\mathbf{k}\uparrow}^{(\lambda)}$, we now define the three quantities
\begin{align*}
S_{x\mathbf{k}}^{(\lambda)} &\equiv \langle\sigma_x\rangle_{\mathbf{k}}^{(\lambda)} = 2u_{\uparrow\mathbf{k}}^{(\lambda)}u_{\downarrow\mathbf{k}}^{(\lambda r)},\\
S_{y\mathbf{k}}^{(\lambda)} &\equiv \langle\sigma_y\rangle_{\mathbf{k}}^{(\lambda)} = 2u_{\uparrow\mathbf{k}}^{(\lambda)}u_{\downarrow\mathbf{k}}^{(\lambda i)},\\
S_{z\mathbf{k}}^{(\lambda)} &\equiv \langle\sigma_z\rangle_{\mathbf{k}}^{(\lambda)} = u_{\uparrow\mathbf{k}}^{(\lambda)2} - u_{\downarrow\mathbf{k}}^{(\lambda r)2} - u_{\downarrow\mathbf{k}}^{(\lambda i)2},
\end{align*}
where $u_{\downarrow\mathbf{k}}^{(\lambda r)}$ and $u_{\downarrow\mathbf{k}}^{(\lambda i)}$ are the real and imaginary parts of $u_{\downarrow\mathbf{k}}^{(\lambda)}$. Together these three expressions defines a unit vector $\mathbf{S}_{\mathbf{k}}^{(\lambda)}$, which at each $\mathbf{k}$ in the Brillouin zone points in the spin direction. Inverting these expressions and using them in Eq.~(\ref{Equation:Berry_connection_two_spinor}), the Berry connection becomes
\begin{equation}
\label{Equation:Berry_connection_four_spinor}
\mathcal{A}_j^{(\lambda)}(\mathbf{k}) = -\frac{1}{2(S_{z\mathbf{k}}^{(\lambda)} + 1)}\!\left(\! S_{x\mathbf{k}}^{(\lambda)}\frac{\partial S_{y\mathbf{k}}^{(\lambda)}}{\partial k_j} - S_{y\mathbf{k}}^{(\lambda)}\frac{\partial S_{x\mathbf{k}}^{(\lambda)}}{\partial k_j}\!\right)\!.
\end{equation}
This expression contains an apparent singularity at $S_{z\mathbf{k}}^{(\lambda)} = -1$, but this happens exactly at those points for which $u_{\uparrow\mathbf{k}}^{(\lambda)} = 0$, points that we have already excluded from the integration domain. These apparent singularities do therefore not present any problem.

It follows from the expression in Eq. (\ref{Equation:Berry_connection_four_spinor}) that the Berry curvature is given by
\begin{equation*}
\mathcal{F}^{(\lambda)}(\mathbf{k}) = -\frac{1}{2(S_{z\mathbf{k}}^{(\lambda)} + 1)^2}\mathbf{T}_{\mathbf{k}}^{(\lambda)}\cdot
\left(\partial_{k_i}\mathbf{S}_{\mathbf{k}}^{(\lambda)}
\times\partial_{k_j}\mathbf{S}_{\mathbf{k}}^{(\lambda)}\right),
\end{equation*}
where $\mathbf{T}_{\mathbf{k}}^{(\lambda)} = (S_{x\mathbf{k}}^{(\lambda)}, S_{y\mathbf{k}}^{(\lambda)}, 2(S_{z\mathbf{k}}^{(\lambda)} + 1))$.
Now, $\mathbf{T}_{\mathbf{k}}^{(\lambda)}$ can be decomposed as $\mathbf{T}_{\mathbf{k}}^{(\lambda)} = (S_{z\mathbf{k}}^{(\lambda)} + 1)^2\mathbf{S}_{\mathbf{k}}^{(\lambda)} + \mathbf{T}_{\perp\mathbf{k}}^{(\lambda)}$, where $\mathbf{T}_{\perp\mathbf{k}}^{(\lambda)}$ is perpendicular to $\mathbf{S}_{\mathbf{k}}^{(\lambda)}$ and therefore also to $\partial_{k_i}\mathbf{S}_{\mathbf{k}}^{(\lambda)}\times
\partial_{k_j}\mathbf{S}_{\mathbf{k}}^{(\lambda)}$, since $\mathbf{S}_{\mathbf{k}}^{(\lambda)}$ is a unit vector.
Thus the Chern number obtained from Eq.~(\ref{Equation:Chern_number}) takes the form
\begin{equation}
\label{Equation:Skyrmion_number}
I^{(\lambda)} = -\frac{1}{4\pi}\int dk_{x}dk_{y}\mathbf{S}_{\mathbf{k}}^{(\lambda)}\cdot
\left(\partial_{k_x}\mathbf{S}_{\mathbf{k}}^{(\lambda)}\times
\partial_{k_y}\mathbf{S}_{\mathbf{k}}^{(\lambda)}\right).
\end{equation}
We note that the integrand in this expression gives zero contribution in the regions where $S_{z\mathbf{k}}^{(\lambda)} = -1$, because either the cross product is zero, or the region on which $S_{z\mathbf{k}}^{(\lambda)} = -1$ has zero measure. It is therefore safe to re-extend the integration back to the full Brillouin zone. The expression in Eq. (\ref{Equation:Skyrmion_number}) is nothing but the winding number for a vector taking values on the unit sphere $S^2$, or because the unit vector represents a spin direction, a spin Skyrmion number. We therefore conclude that when $|\Psi^{(\lambda)}(\mathbf{k})\rangle$ is a two-component spinor field defined over the 2D Brillouin zone, the Chern number is a Skyrmion number.

\section{Ferromagnetic semiconductor-superconductor heterostructures}
\label{Section:Skyrmion_in_topological_superconductor}
\subsection{Skyrmion number for 2D systems}
Now we turn our attention to the Chern number that is used for the classification of 2D semiconductor heterostructures which hosts Majorana fermions in the topologically non-trivial phase.\cite{Sato09, Sato10,Sau10} More precisely, we consider here the Chern number for four-component spinor fields which arise as solutions to the eigenvalue equation for Hamiltonians of the form
\begin{equation}
\label{Equation:Hamiltonian}
\mathcal{H}(\mathbf{k}) = \left[\begin{array}{cc}
H_0(\mathbf{k}) & \Delta(\mathbf{k})\\
\Delta^{\dagger}(\mathbf{k}) & - H_0^{T}(-\mathbf{k})
\end{array}\right],
\end{equation}
in two spatial dimensions, which is written in the basis $|\Psi^{(\lambda)}(\mathbf{k})\rangle = \left[\begin{array}{cc}\mathbf{u}_{\mathbf{k}}^{(\lambda)T} & \mathbf{v}_{-\mathbf{k}}^{(\lambda)T}\end{array}\right]^T = \left[\begin{array}{cccc}u_{\uparrow\mathbf{k}}^{(\lambda)} & u_{\mathbf{\downarrow\mathbf{k}}}^ {(\lambda)} & v_{\uparrow-\mathbf{k}}^ {(\lambda)} & v_{\downarrow-\mathbf{k}}^ {(\lambda)}\end{array}\right]^T$. Here $\mathbf{u}_{\mathbf{k}}^{(\lambda)}$ and $\mathbf{v}_{\mathbf{k}}^{(\lambda)}$ are the electron and hole components of the eigenstates. Eq.~(\ref{Equation:Hamiltonian}) is the most general Bogoliubov-de Gennes Hamiltonian, and a ferromagnetic semiconductor-superconductor heterostructure can be modelled by setting\cite{Sato10, Bjornson13}
\begin{align}
\label{Equation:Hamiltonian_specific}
H_0(\mathbf{k}) =& \epsilon(\mathbf{k}) - V_z\sigma_z - \mathcal{L}_0(\mathbf{k})\cdot\boldsymbol{\sigma},\\
\Delta(\mathbf{k}) =& i\Delta\sigma_y.
\end{align}
Here $\epsilon(\mathbf{k})$ is the band structure of a lightly doped semiconductor which we approximate by $\epsilon(\mathbf{k}) = -2t(\cos(k_x) + \cos(k_y)) - \mu$ with $\mu \approx \pm 4t$, which is the simplest possible band structure on a square lattice. Furthermore, $\mathcal{L}_0(\mathbf{k}) = \alpha(\sin(k_y), -\sin(k_x), 0)$ provides a Rashba spin-orbit coupling and $V_z$ is a Zeeman exchange field. Finally $\Delta$ provides an $s$-wave superconducting term induced through proximity effect to a conventional superconductor.

Because the combination of spin-orbit interaction and Zeeman term forces the Bogoliubov-de Gennes Hamiltonian to be written on a four-by-four form, the degrees of freedom has been artificially doubled. We therefore expect the information of the system, such as the topological invariant, to be redundantly encoded in the resulting spinors. The purpose here is to show that it is possible to use only half of this information to calculate the topological invariant. Especially we will show that this construction leads us to an interpretation of also this Chern number as a Skyrmion number.

The first step in this process lies in realizing that because the order parameter is nowhere zero, $\Delta$ can be inverted, which together with the two first lines of the eigenvalue equation $\mathcal{H}|\Psi^{(\lambda)}\rangle = E^{(\lambda)}|\Psi^{(\lambda)}\rangle$ results in
\begin{equation}
\label{Equation:v_of_u}
\mathbf{v}_{-\mathbf{k}}^{(\lambda)} = \Delta^ {-1}(\mathbf{k})\left(E^{(\lambda)} - H_0(\mathbf{k})\right)\mathbf{u}_{\mathbf{k}}^{(\lambda)}.
\end{equation}
This expression, together with the normalization condition $\langle\Psi^{(\lambda)}|\Psi^{(\lambda)}\rangle = 1$, implies that $\mathbf{u}_{\mathbf{k}}^{(\lambda)}$ and $\mathbf{v}_{-\mathbf{k}}^{(\lambda)}$ is always finite. The importance of this equation is that it shows that the electron components alone contains all the information about the system. Further it allows us to define
\begin{align}
\label{Equation:Renormalization}
\mathbf{\bar{u}}_{\mathbf{k}}^{(\lambda)} = \frac{\mathbf{u}_{\mathbf{k}}^{(\lambda)}}{\langle\mathbf{u}_{\mathbf{k}}^{(\lambda)}|\mathbf{u}_{\mathbf{k}}^{(\lambda)}\rangle},
\end{align}
the renormalized spin component. It follows from the discussion in the previous section that, because $\mathbf{\bar{u}}$ is a normalized two-component spinor, it is possible to define a Skyrmion number
\begin{align}
\label{Equation:Skyrmion_number_four_component}
I_{\mathbf{\bar{u}}}^{(\lambda)} =& -\frac{1}{4\pi}\int dk_xd_ky\mathbf{S}_{\mathbf{k}}^{(\lambda)}\cdot\left(
\partial_{k_x}\mathbf{S}_{\mathbf{k}}^{(\lambda)}\times
\partial_{k_y}\mathbf{S}_{\mathbf{k}}^{(\lambda)}\right),\\
\label{Equation:Renormalized_spin}
\mathbf{S}_{\mathbf{k}}^{(\lambda)} =& \mathbf{\bar{u}}_{\mathbf{k}}^{(\lambda)\dagger}\boldsymbol{\sigma}
\mathbf{\bar{u}}_{\mathbf{k}}^{(\lambda)}.
\end{align}
This is clearly a topological invariant, which can only change when two bands cross each other.\footnote{The reason it can change at the band crossing is because the assignment of labels $\lambda$ to different states becomes arbitrary at the degenerate points.}

\begin{figure}
\includegraphics[width=245pt]{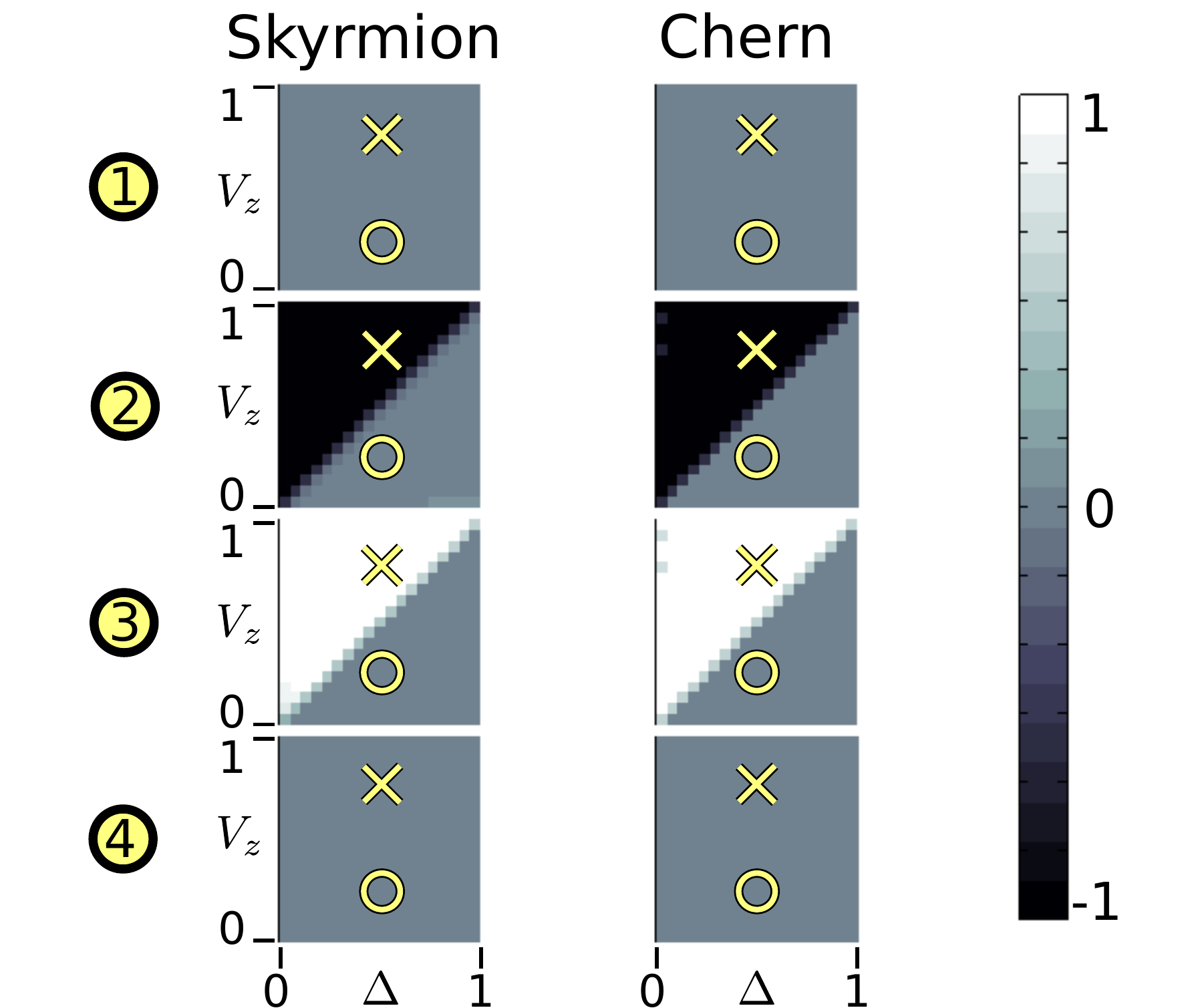}
\caption{(Color online) Skyrmion and Chern numbers for each of the four bands for different parameters $V_z$ and $\Delta$ for a 2D ferromagnetic and Rashba spin-orbit coupled semiconductor-superconductor heterostructure. Band 1 and 4 are topologically trivial for all $V_z$ and $\Delta$, while band 2 and 3 acquire a topologically non-trivial configuration for $V_z > \Delta$. X and O mark the values for which spin configurations are shown in Fig.~\ref{Figure:Spin_configurations}.}
\label{Figure:Skyrmion_vs_chern}
\end{figure}

We now show that the Skyrmion number in Eq. (\ref{Equation:Skyrmion_number_four_component}) is the same as the Chern number calculated from Eq. (\ref{Equation:Chern_number}-\ref{Equation:Berry_connection}) for the full four-component spinor. For this purpose we plot, for multiple parameter values $V_z$ and $\Delta$, both the Chern number and Skyrmion number side by side for all four bands in Fig. \ref{Figure:Skyrmion_vs_chern}. This is done by numerically diagonalizing Eq. (\ref{Equation:Hamiltonian}) and using Eq. (\ref{Equation:Skyrmion_number_four_component}) to calculate the Skyrmion number. The Chern number is in principle given by Eq. (\ref{Equation:Chern_number}-\ref{Equation:Berry_connection}), but the well known alternative expression\cite{TopologicalInsulatorsAndTopologicalSuperconductors}
\begin{widetext}
\begin{equation*}
\label{Equation:Numerical_chern_number}
I^{(\lambda)} = -\frac{1}{2\pi}\int dk_xdk_yIm\left(\sum_{\rho\neq \lambda}\epsilon^{ij}\frac{\langle\Psi^{(\lambda)}|\partial_{k_i}\mathcal{H}|\Psi^{(\rho)}\rangle\langle\Psi^{(\rho)}|\partial_{k_j}\mathcal{H}|\Psi^{(\lambda)}\rangle}{(E^{(\rho)} - E^{(\lambda)})^2}\right)
\end{equation*}
\end{widetext}
has instead been used for numerical reasons. Fig. \ref{Figure:Skyrmion_vs_chern} shows that the two expressions clearly agree for all plotted values.
One might here object that the two expressions have only been shown to agree for certain parameters. However, we already know that both expressions are topological invariants, which can only change when bands cross each other. If they agree at a single point in one topological phase, they must therefore agree at all points in the same phase. The only possibility to find another set of parameters where they do not agree would be if the system passed through a topological phase transition and the change in topological number was different across the transition for the two expressions. However, we clearly see that they agree on both sides of the topological phase transition in Fig \ref{Figure:Skyrmion_vs_chern}. Further, this is the only phase transition of interest in realistic materials involving semiconductors,\cite{Bjornson13} we therefore conclude that it is safe to use the Skyrmion expression instead of the full four component Chern number.

\subsubsection{Experimental implications}
\begin{figure*}
\includegraphics[width=465pt]{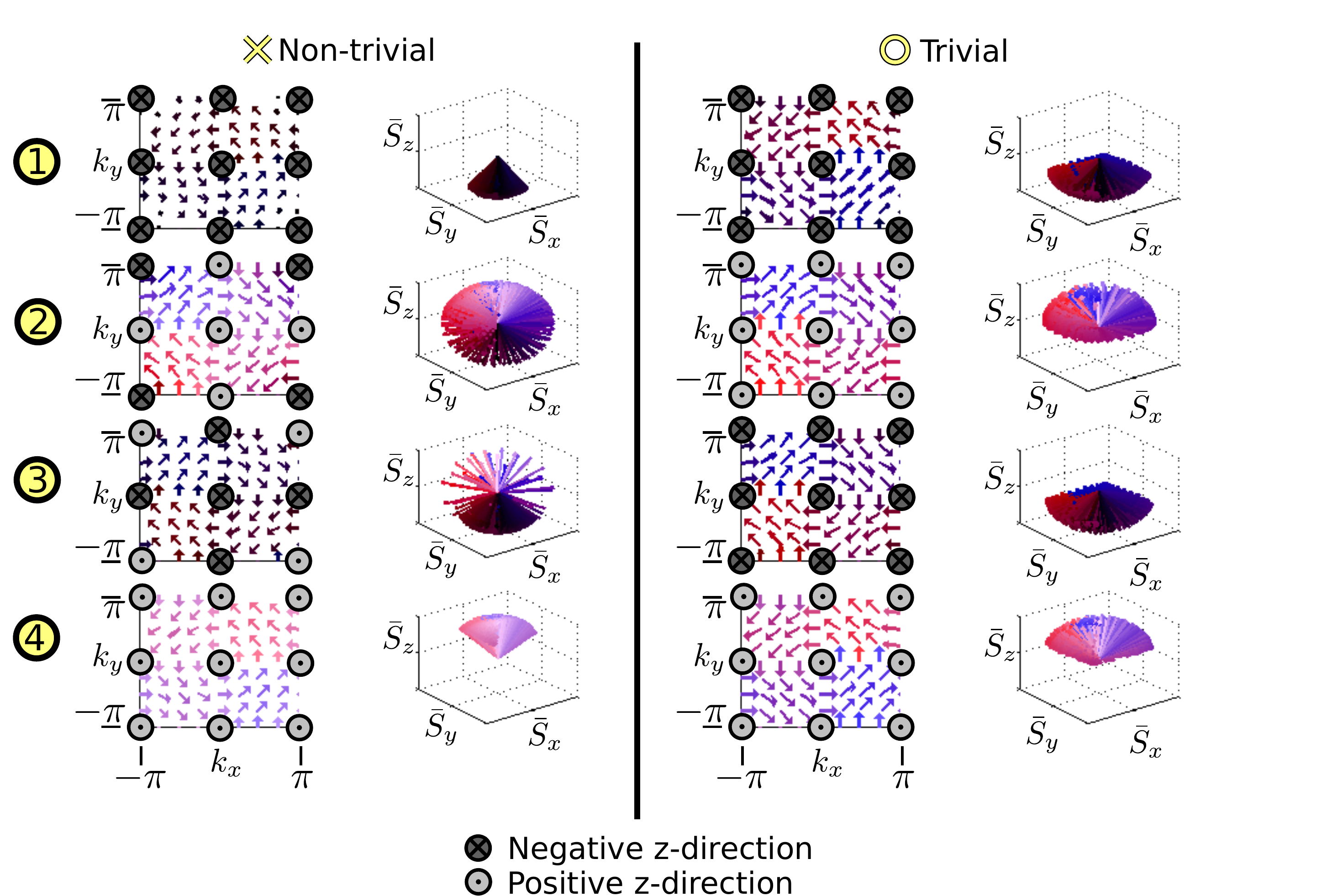}
\caption{(Color online) Spin configurations for the four bands at the topologically non-trivial point X  and trivial point O in Fig.~\ref{Figure:Skyrmion_vs_chern}. In the left columns the spins are plotted in the Brillouin zone, while in the right columns all spins are plotted with a common origin. Dark/bright colors and arrow notation indicate negative/positive $z$-component, while red/blue indicates azimuthal angle. Band 1 and 4 can be seen to have a trivial spin structure (continuously deformable to be aligned with the $\pm z$-axis), while band 2 and 3 develops a Skyrmion structure in the non-trivial phase.}
\label{Figure:Spin_configurations}
\end{figure*}

The main advantage of the Skyrmion number formulation over the more abstract Chern number is that Eq.~(\ref{Equation:Skyrmion_number_four_component}) has a direct physical significance: it is the winding of the spin direction for the electrons at different points in the Brilloiun zone. This allows for the topological invariant to be directly extracted using any method where spin-resolved band structure data is available. In fact, Eq.~(\ref{Equation:Skyrmion_number_four_component}) itself becomes nearly obsolete once spin-polarized band structure data is obtained, as the topological invariant can be deduced from simple inspection, as seen in Fig. \ref{Figure:Spin_configurations}. Here we have plotted the spin configurations both in the trivial and non-trivial phase for the specific parameter values marked by $X$ and $O$ in Fig. \ref{Figure:Skyrmion_vs_chern}. As clearly visible, Skyrmions appears in the two bands closest to the band crossing. We here also mention that Eq. (\ref{Equation:Berry_connection}) is antisymmetric with respect to the particle-hole symmetry, and the Skyrmion numbers in the particle-hole related bands are therefore also related by a minus sign.

\subsection{1D wires}
The formalism developed above for 2D semiconductor heterostructures can be expanded to 1D wires. This is particularly interesting as the so far most promising system for realization of Majorana fermions are 1D wires consisting of a similar combination of ferromagnet-semiconductor-superconductor components. This system can be described by using the Hamiltonian
\begin{align}
\label{Equation:One_dimensional}
H_{0}^{1D}(\mathbf{k}) = -2t\cos(k_x) - \tilde{\mu} - V_z\sigma_z + \alpha\sin(k_x)\sigma_y
\end{align}
in Eq. (\ref{Equation:Hamiltonian}), and is up to choice of $\sigma$-matrix labels the lattice equivalent of previously studied continuum models.\cite{Lutchyn10, Oreg10}
We arrive at this expression through dimensional reduction of Eq. (\ref{Equation:Hamiltonian_specific}) by setting $\tilde{\mu} = \mu - 2t$ and $k_y = \pi$. Thus the spin structure of the 1D model follows from that along $k_y = \pi$ in Fig. \ref{Figure:Spin_configurations}. In addition to the 1D case being of large interest in itself, a detailed study of it also gives a better understanding of the origin of the Skyrmions in 2D.

\begin{figure*}
\includegraphics[width=465pt]{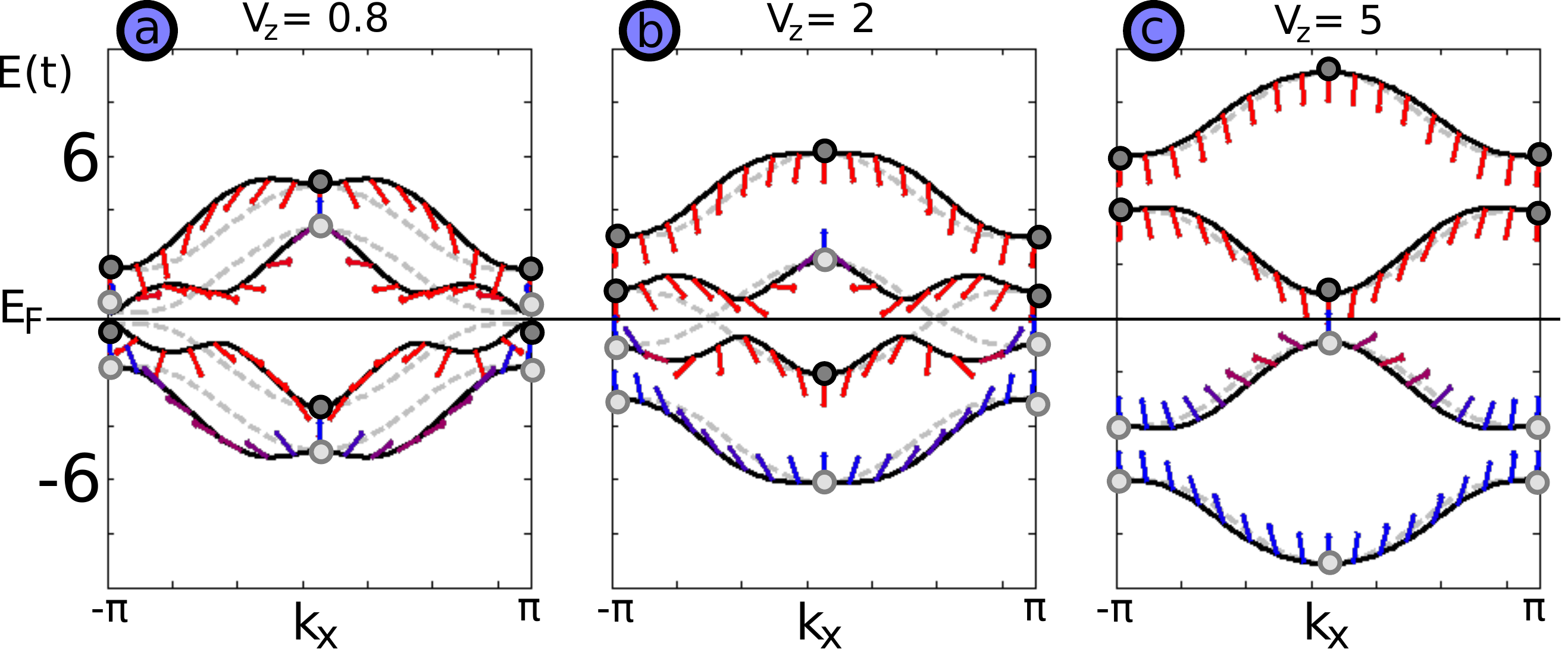}
\caption{(Color online) Band structure of the Hamiltonian in Eq. (\ref{Equation:One_dimensional}) for $t = 1, \Delta = 1, \tilde{\mu} = 2, \alpha = 2$ and three representative values of $V_z$. The topological structure being (a) trivial , (b) non-trivial, (c) trivial. Arrows indicates how the spins twists in the $y,z$-plane. In the background the same band structure for $\alpha = 0$ has been plotted (dashed lines). For $\alpha = 0$, the spins in each band are either parallel or anti-parallel with the Zeeman term. When $\alpha$ is turned on, this remains true at $k_x = 0,\pm\pi$, where $\alpha$ has no effect, but elsewhere both the band and spin structure is modified. Up(down) spins at the high symmetry points has been highlighted with bright(dark) spots. Note how in the intermediate case (b), the down spin is below the Fermi level at $k_x = 0$, while an up spin is below the Fermi level at $k_x = \pi$.
%
%
}
\label{Figure:One_dimensional_band_structure}
\end{figure*}

In Fig. \ref{Figure:One_dimensional_band_structure} we plot the band structure for the 1D wire in Eq. (\ref{Equation:One_dimensional}) for representative values of the different topological phases, as well as the band structure for $\alpha = 0$. On top of these bands the renormalized spins given by Eq. (\ref{Equation:Renormalized_spin}) are also indicated. For $\alpha = 0$ all spins in any specific band is either parallel or anti-parallel with the Zeeman term $V_z$. As $V_z$ is turned up, the $\alpha = 0$ bands with up spins are pushed down in energy, while the down spin bands are pushed up. In the intermediate parameter region (b), two bands with opposite spin overlap each other. Because these two bands bend in opposite directions, a situation is created where at some points in the Brillouin zone down spins are below the Fermi level, while at other points instead up spins are below.

Once $\alpha$ is turned on, the spin structure does not change at $k_x = 0,\pm\pi$ where the spin-orbit interaction has no effect. Most importantly, in the topologically non-trivial region (b), the degeneracy between the two $\alpha = 0$ bands at the crossing points at the Fermi level is lifted. This results in two separate bands around the Fermi level, which have different spin direction at $k_x = 0$ and $k_x = \pm\pi$. Away from these high symmetry points the $\alpha$ term twists the spin into the $y$-direction, thereby creating a smooth transition from the up(down) spin at $k_x = 0$, to the down(up) spin at $k_x = \pm\pi$ in the upper(lower) band.

Finally imagine turning up $V_z$ even further before turning on $\alpha$.
Although self-consistent calculations has shown that this is a rather unphysical situation, since $V_z$ destroys superconductivity,\cite{Bjornson13} this case is useful to consider for conceptual purposes. Eventually we reach the situation where the $\alpha = 0$ bands no longer overlap, such that all up spins are below the Fermi level while all down spins are above. This is represented by the band structure in Fig. \ref{Figure:One_dimensional_band_structure}(c). In this case the spins at $k_x = 0$ and $k_x = \pm\pi$ are the same in each band and the only effect of turning on $\alpha$ is to somewhat deform the spin structure. This deformation no longer twists the spin around a full circle as we move along $-\pi\rightarrow 0\rightarrow\pi$ in $k_x$-space, but rather just twists it toward the $y$-axis and then back toward the same $z$-direction again. It is in this sense most useful to think of the topologically non-trivial phase as the parameter interval where the spin at one, but only one, of the points $k_x = 0$ and $k_y = \pm\pi$ has been interchanged between the two bands closest to the Fermi energy, thereby forcing the spin to twist from up to down and back to up again as we trace out the full 1D Brillouin zone. This is the very origin of the twisted spin structure.

\begin{figure}
\includegraphics[width=245pt]{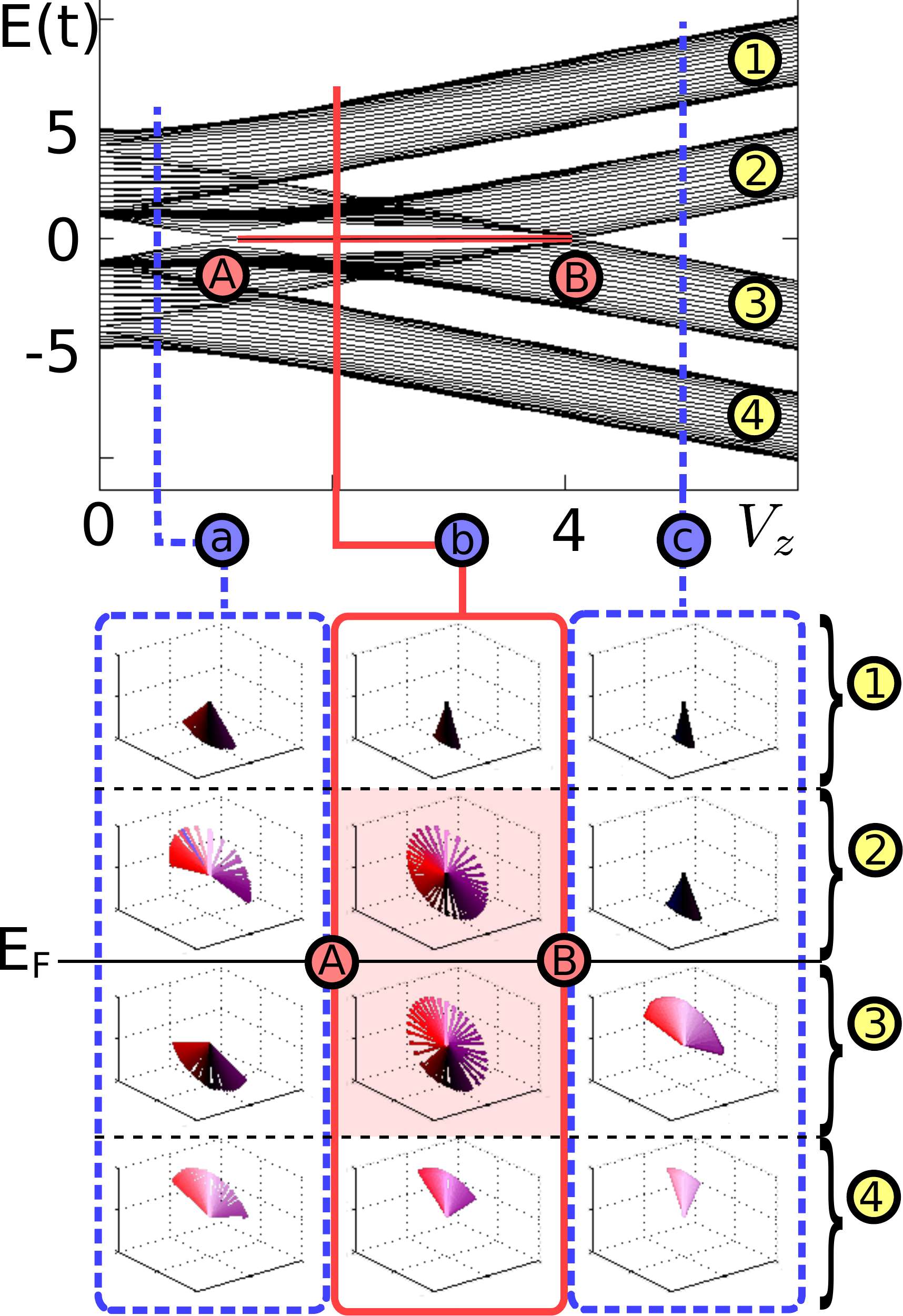}
\caption{(Color online) (Top) Energy spectrum as a function of $V_z$ for a 1D wire with open boundary conditions. Topologically non-trivial region $V_z \in [1, 4]$ hosts Majorana fermions, seen as horizontal lines at $E = 0$. Interchange of spins between the bands at $k_x = \pm\pi$ and $k_x = 0$, marking the entrance/exit of the topologically non-trivial phase, occurs at A and B.
(Bottom) Spin structure in the four bands for the same three representative values of $V_z$ as in Fig. \ref{Figure:One_dimensional_band_structure}, and as indicated by the vertical lines at the top. Inside the non-trivial phase the spin structure in the two bands closest to the Fermi energy covers a whole circle (shaded red).}
\label{Figure:One_dimensional}
\end{figure}

In Fig. \ref{Figure:One_dimensional} we provide further information which link the spin structures, band crossings, and existence of Majorana fermions to each other. In the energy spectrum (top) we clearly see that the region between ($A$) and ($B$) hosts the characteristic Majorana zero mode. Comparing to the band structures in Fig. \ref{Figure:One_dimensional_band_structure}, the band structure in \ref{Figure:One_dimensional_band_structure}(a) transforms into the band structure \ref{Figure:One_dimensional_band_structure}(b) at the point which is marked by ($A$) in Fig. \ref{Figure:One_dimensional}, while the transition from the band structure in \ref{Figure:One_dimensional_band_structure}(b) to that in \ref{Figure:One_dimensional_band_structure}(c) happens at ($B$). In other words, ($A$) is the point where the two bands closest to the Fermi level meet at $k_x = \pm\pi$, while ($B$) is the point where they meet at $k_x = 0$. Thus the Majorana fermion appears in the intermediate region where one, but not both, of the spins at the high symmetry points has interchanged band with each other. In Fig. \ref{Figure:One_dimensional}(bottom) we also clearly see that the spin structure forms a circle in this and only this intermediate region (shaded red).
%

From these observations it is clear how the one-dimensional analogue of Eq. (\ref{Equation:Skyrmion_number_four_component}) is to be defined, namely
\begin{align}
I_{\mathbf{\bar{u}}}^{(\lambda)} =& \frac{1}{2\pi}\int dk_x \mathbf{\hat{x}}\cdot\left(\mathbf{S}_{\mathbf{k}}^{(\lambda)}
\times\partial_{k_x}\mathbf{S}_{\mathbf{k}}^{(\lambda)}\right).
\end{align}
In the case a different combination of $\sigma$-matrices are used in the Hamiltonian in Eq. (\ref{Equation:One_dimensional}), the unit vector $\mathbf{\hat{x}}$ should be replaced by the unit vector which corresponds to the $\sigma$-matrix which is not used in the Hamiltonian. To understand why this is the correct expression for the topological invariant, it is enough to realize that the expression inside the parenthesis is a vector pointing in the $\mathbf{\hat{x}}$-direction, with magnitude equal to the rate of change of $\mathbf{S}_{\mathbf{k}}^{(\lambda)}$. Also note that although the expression in principle is able to calculate an arbitrary winding of $\mathbf{S}_{\mathbf{k}}^ {(\lambda)}$, in reality $\mathbf{S}_{\mathbf{k}}^{(\lambda)}$ will only give rise to $0, \pm 1$, because the spin orbit interaction is only able to twist the spin once across the 1D Brillouin zone. The sign is related to whether the spins are twisted clockwise or counter-clockwise, and is changed if $V_z \rightarrow -V_z$ or $\alpha \rightarrow -\alpha$, but is of no importance as far as the existence of Majorana fermions is considered. That is, the topological invariant is effectively a $\mathbb{Z}_2$ invariant.

\subsection{2D case revisited}
Above we showed how the topologically non-trivial spin structure arises in the 1D case when the spins at one, but not both, of the high symmetry points $k_x = 0,\pm\pi$ interchange band with each other. Therefore making the spin at $k_x = 0$ and $k_x = \pm\pi$  within the same band point in opposite directions. It is now worthwhile to revisit the 2D case. A quick look at Fig. \ref{Figure:Spin_configurations} reveals that the observed Skyrmion arise in a similar way; it arise when the spin at different high symmetry points in the same band point in different directions. Also in this case it is the spin-orbit interaction which provides the smooth transition between these conflicting spin directions in the region between the high symmetry points, thereby creating a Skyrmion. This is a very powerful observation from an experimental point of view, because this allows for identification of the topological phase to be performed by mapping out the spin direction only at the high symmetry points. If the system has a full energy gap, and any spin at a high symmetry point not agrees with the spin direction at another high symmetry point in the same band, then a Skyrmion necessarily have developed in that band.

\section{Discussion}
In summary we have shown that the topological phase of 2D ferromagnetic semiconductor-superconductor heterostructures, previously classified using Chern numbers, and their 1D analogues, most easily can be thought of as phases hosting band structure Skyrmions. The Skyrmion topological invariant is the sum of the number of momentum-space Skyrmions in the spin configuration in the occupied bands.

This result shows that any technique capable of extracting the spin-resolved band structure can be used to determine the topological phase of the system. Moreover, it is enough to map out the spin direction only at the high symmetry points, if in addition the system is known to be gapped throughout the Brillouin zone. In the case some spin direction at some high symmetry point do not agree with the spin direction at another high symmetry point in the same band, then a Skyrmion has developed in the band, and the system is in a topologically non-trivial phase. Thus for example ARPES seems ideally suited to directly measure the topological invariant.

Indicative information is also available in the topologically trivial phase by studying the spreading of the spins in the full Brillouin zone. If the spins are observed to cover an ever greater area of the unit sphere as the bands approach each other towards a band crossing across the energy gap, then that is a clear indication of the band crossing being a topological phase transition. This could prove especially useful in cases where different band structures are engineered but a potential topological phase transition is initially difficult to obtain.

Our treatment of the 1D case also makes it clear that the 1D wires and 2D heterostructures can be treated within the same framework. The only difference between the 1D and 2D case is that the spins in the 1D case traces out a circle instead of a full Skyrmion. We think that this unified framework, and the intuitively more accessible Skyrmion number, will prove helpful in both experimental investigation of these materials, as well as in theoretical model building. Due to the physical accessibility of the spin configuration by a spin-resolved band structure imaging techniques, be it experimental or numerical, this result should significantly accelerate the search for topologically non-trivial superconducting states.

\section{Acknowledgements}
We are grateful to J.~C.~Budich for discussions and the Swedish Research Council (VR) for financial support.

\end{document}